\documentclass[12pt]{article}
\usepackage{amsmath,amssymb,amsthm,amsxtra,overpic,bbm,bm,epsfig,subfigure}
\usepackage{hyperref}
\usepackage{mathrsfs}
\usepackage{enumitem}
\usepackage{graphicx}
\usepackage{color}
\usepackage{comment}
\usepackage{epstopdf}
\usepackage{float}
\usepackage{cite}
\usepackage{slashed,stmaryrd}

\def\thefootnote{\fnsymbol{footnote}}

\begin{document}

\vspace{0.2cm}

\begin{center}
{\Large\bf The Smallest Neutrino Mass Revisited}
\end{center}

\vspace{0.2cm}

\begin{center}
{\bf Shun Zhou}~\footnote{E-mail: zhoush@ihep.ac.cn}
\\
\vspace{0.2cm}
{\small
Institute of High Energy Physics, Chinese Academy of Sciences, Beijing 100049, China\\
School of Physical Sciences, University of Chinese Academy of Sciences, Beijing 100049, China\\}
\end{center}

\vspace{1.5cm}

\begin{abstract}
As is well known, the smallest neutrino mass turns out to be vanishing in the minimal seesaw model, since the effective neutrino mass matrix $M^{}_\nu$ is of rank two due to the fact that only two heavy right-handed neutrinos are introduced. In this paper, we point out that the one-loop matching condition for the effective dimension-five neutrino mass operator can make an important contribution to the smallest neutrino mass. By using the available one-loop matching condition and two-loop renormalization group equations in the supersymmetric version of the minimal seesaw model, we explicitly calculate the smallest neutrino mass in the case of normal neutrino mass ordering and find $m^{}_1 \in [10^{-10}, 10^{-8}]~{\rm eV}$ at the Fermi scale $\Lambda^{}_{\rm F} = 91.2~{\rm GeV}$, where the range of $m^{}_1$ results from the uncertainties on the choice of the seesaw scale $\Lambda^{}_{\rm SS}$ and on the input values of relevant parameters at $\Lambda^{}_{\rm SS}$.
\end{abstract}

%\begin{flushleft}
%\hspace{0.8cm} PACS number(s):
%\end{flushleft}

\def\thefootnote{\arabic{footnote}}
\setcounter{footnote}{0}

\newpage

\section{Introduction}\label{sec:intro}

Although neutrino oscillation experiments have firmly established that neutrinos are massive, the smallest neutrino mass has not yet been determined experimentally~\cite{Xing:2019vks, Capozzi:2017ipn}. In the case of normal neutrino mass ordering (NO) with $m^{}_1 < m^{}_2 < m^{}_3$, the latest global-fit analysis of neutrino oscillation data~\cite{Esteban:2020cvm, Nufit} yields two neutrino mass-squared differences $\Delta m^2_{21} \equiv m^2_2 - m^2_1 \approx 7.42\times 10^{-5}~{\rm eV}^2$ and $\Delta m^2_{31} \equiv m^2_3 - m^2_1 \approx 2.514\times 10^{-3}~{\rm eV}^2$, so one can immediately obtain $m^{}_2 = \sqrt{\Delta m^2_{21}} \approx 8.61~{\rm meV}$ and $m^{}_3 = \sqrt{\Delta m^2_{31}} \approx 50.1~{\rm meV}$ for $m^{}_1 = 0$. In the case of inverted neutrino mass ordering (IO) with $m^{}_3 < m^{}_1 < m^{}_2$, we have $\Delta m^2_{32} \equiv m^2_3 - m^2_2 \approx -2.497\times 10^{-3}~{\rm eV}^2$ and $\Delta m^2_{21} \approx 7.42\times 10^{-5}~{\rm eV}^2$~\cite{Esteban:2020cvm}, and thus $m^{}_2 = \sqrt{-\Delta m^2_{32}} \approx 50.0~{\rm meV}$ and $m^{}_1 = \sqrt{- \Delta m^2_{32} - \Delta m^2_{21}} \approx 49.2~{\rm meV}$ for $m^{}_3 = 0$. Either $m^{}_1 = 0$ in the NO case or $m^{}_3 = 0$ in the IO case is still compatible with neutrino oscillation data, as well as other observational constraints on the absolute neutrino mass from tritium beta decays, neutrinoless double-beta decays and cosmology~\cite{Capozzi:2017ipn}.

On the theoretical side, it is interesting to ask whether the vanishing neutrino mass $m^{}_1 = 0$ or $m^{}_3 = 0$ can be naturally realized in any neutrino mass models. At the leading order, the answer is affirmative in the so-called minimal type-I seesaw model (MSM), which extends the Standard Model (SM) with only two right-handed neutrino singlets $N^{}_{i{\rm R}}$ (for $i = 1, 2$). More explicitly, the gauge-invariant Lagrangian for the MSM is given by~\cite{Kleppe:1995zz, Ma:1998zg, King:1999mb, King:2002nf, Frampton:2002qc, Guo:2006qa, Xing:2020ald}
\begin{eqnarray}
{\cal L}^{}_{\rm MSM} = {\cal L}^{}_{\rm SM} + \overline{N^{}_{i {\rm R}}} {\rm i}\slashed{\partial} N^{}_{i {\rm R}} - \left[ \overline{\ell^{}_{\alpha {\rm L}}} \left(Y^{}_\nu\right)^{}_{\alpha i} \widetilde{H} N^{}_{i{\rm R}} + \frac{1}{2} \overline{N^{\rm C}_{i{\rm R}}} \left(M^{}_{\rm R}\right)^{}_{ij} N^{}_{j{\rm R}} + {\rm h.c.}\right] \; ,
%     (1)
\label{eq:Lag}
\end{eqnarray}
where $\ell^{}_{\alpha {\rm L}}$ (for $\alpha =e, \mu, \tau$) stand for three left-handed lepton doublets, $\widetilde{H} \equiv i\sigma^{}_2 H^*$ with $H \equiv \left(\varphi^+, \varphi^0\right)^{\rm T}$ for the Higgs doublet, $Y^{}_\nu$ for the $3\times 2$ Yukawa coupling matrix, and $M^{}_{\rm R}$ for the $2\times 2$ Majorana mass matrix of right-handed neutrinos. At the low-energy scale, the effective neutrino mass operator can be obtained by integrating out two heavy Majorana neutrinos~\cite{Weinberg:1979sa}, i.e.,
\begin{eqnarray}
{\cal O}^{}_\nu = - \frac{1}{2} \kappa^{}_{\alpha \beta} \overline{\ell^{}_{\alpha {\rm L}}} \widetilde{H} \widetilde{H}^{\rm T} \ell^{\rm C}_{\beta {\rm L}} + {\rm h.c.} \; ,
%     (2)
\label{eq:Weinberg}
\end{eqnarray}
where $\kappa = Y^{}_\nu M^{\rm -1}_{\rm R} Y^{\rm T}_\nu$ is in general a $3\times 3$ complex and symmetric matrix. Therefore, the effective Majorana mass matrix of three light neutrinos is given by $M^{}_\nu = \kappa v^2$ with $v \equiv \langle \varphi^0\rangle \approx 174~{\rm GeV}$ being the vacuum expectation value (vev) of the Higgs field. As the Majorana mass matrix $M^{}_{\rm R}$ is of rank two, one reaches the conclusion that $M^{}_\nu$ is of rank two as well and the smallest neutrino mass is vanishing (i.e., $m^{}_1 = 0$ for NO or $m^{}_3 = 0$ for IO).

However, as pointed out in Ref.~\cite{Davidson:2006tg}, if the two-loop renormalization-group (RG) running effects of the effective neutrino mass parameter $\kappa$ are taken into account, then the rank of $M^{}_\nu$ will be increased from two to three. Consequently, the smallest neutrino mass in the MSM has been found to be $m^{}_{\rm L} \sim 10^{-13}~{\rm eV}$ in the SM framework, while $m^{}_{\rm L} \sim 10^{-10}~{\rm eV} \cdot (\tan \beta/10)^4$ in the Minimal Supersymmetric Standard Model (MSSM), where $\tan \beta \equiv v^{}_{\rm u}/v^{}_{\rm d}$ with $v^{}_{\rm u} = v \sin \beta$ and $v^{}_{\rm d} = v \cos\beta$ being the vev's of two Higgs doublets $H^{}_{\rm u}$ and $H^{}_{\rm d}$, respectively. In the SM, the two-loop RG running of neutrino masses, flavor mixing angles and CP-violating phases has been studied carefully by the authors of Ref.~\cite{Xing:2020ezi}. In this connection, it is interesting to notice that a lower bound on the smallest leptonic mixing angle can also be derived~\cite{Ray:2010fa}. Notice also that the smallest neutrino mass is denoted here by $m^{}_{\rm L}$, which is actually $m^{}_1$ in the NO case or $m^{}_3$ in the IO case. In the present work, we stress that the one-loop matching condition at the seesaw scale $\Lambda^{}_{\rm SS} \sim {\cal O}(M^{}_{\rm R})$ between the full MSM and the effective theory should be implemented together with the two-loop RG running for self-consistency. If this is correctly done, the impact of one-loop matching on the smallest neutrino mass will in general be more significant than that of two-loop RG running. We explicitly demonstrate such an observation in the MSSM, for which both one-loop matching conditions~\cite{Antusch:2015pda, CandiadaSilva:2020hxj} and two-loop RG equations~\cite{Antusch:2002ek} are available in the type-I seesaw model and the corresponding effective theories after the decoupling of heavy Majorana neutrinos.

The remaining part of this work is organized as follows. In Sec.~\ref{sec:framework}, we briefly introduce the supersymmetric version of the MSM and summarize the one-loop matching conditions and two-loop RG equations derived in Refs.~\cite{Antusch:2015pda, Antusch:2002ek} in order to establish our notations. The impact of the one-loop matching on the smallest neutrino mass is discussed in Sec.~\ref{sec:mass}, where the RG running effects are examined as well. Finally, we conclude in Sec.~\ref{sec:sum}.

\section{One-loop Matching}\label{sec:framework}

To scrutinize the type-I seesaw model~\cite{Minkowski:1977sc} at a superhigh-energy scale with precision measurements at low energies, one may follow the approach of effective field theories~\cite{Weinberg:1978ym}. The basic strategy is as follows. First, we change to the mass basis of heavy Majorana neutrinos $N^{}_i$ (for $i = 1, 2$), where $M^{}_{\rm R} = {\rm Diag}\{M^{}_1, M^{}_2\}$ is diagonal with $M^{}_1 \lesssim M^{}_2$, in the full theory and construct an effective theory by integrating out the heaviest particle $N^{}_2$. Second, one matches the full theory into the effective theory at the energy scale $\mu = M^{}_2$ and finds out the matching conditions of physical parameters. Third, starting with the initial values of physical parameters in the effective theory at the matching scale, one then implements the RG equations to run those parameters to the next threshold of the heavy particle $N^{}_1$. Such a procedure can be continued until the relevant energy scale of low-energy observations is reached, such as the Fermi scale $\Lambda^{}_{\rm F} \equiv M^{}_Z = 91.2~{\rm GeV}$. The matching and running of physical parameters in the type-I seesaw model have been extensively studied in the literature~\cite{Chankowski:1993tx, Babu:1993qv, Antusch:2001ck, Antusch:2001vn, Casas:1999tg, Antusch:2003kp, Mei:2003gn, Antusch:2005gp, Mei:2005qp, Ohlsson:2013xva}

Now we apply the aforementioned strategy to the MSM in the MSSM framework.\footnote{The complete set of one-loop matching conditions for the neutrino mass operator in the type-I seesaw model has recently been derived in Ref.~\cite{Zhang:2021jdf}, but the two-loop RG equations in the associated effective theories are still lacking. Once they are known, one can immediately apply the same procedure presented in this work to perform a thorough study of the smallest neutrino mass from one-loop matching and two-loop RG running in the MSM in the SM framework. Some interesting results relevant for the one-loop matching conditions in the seesaw models have previously been found in Refs.~\cite{Pilaftsis:1991ug, Dev:2012sg}.} In this case, the superpotential relevant for lepton masses and flavor mixing can be written as
\begin{eqnarray}
{\cal W}^{}_l = \left(Y^*_l\right)^{}_{\alpha \beta} \widehat{L^{}_\alpha} \widehat{H}^{}_{\rm d} \widehat{E}^{\rm C}_\beta +  \left(Y^*_\nu\right)^{}_{\alpha i} \widehat{L^{}_\alpha} \widehat{H}^{}_{\rm u} \widehat{N}^{\rm C}_i + \frac{1}{2} \left(M^*_{\rm R}\right)^{}_{ij} \widehat{N}^{\rm C}_i  \widehat{N}^{\rm C}_j \; ,
%     (3)
\label{eq:super}
\end{eqnarray}
where the chiral superfields are indicated by hats and the summation over the flavor indices is implied. It is worth mentioning that the Yukawa coupling and mass matrices in the superpotential have been taken to be their complex conjugates in order to comply with our conventions in the SM. To simplify the matching and running procedure, we further assume that the masses of two heavy Majorana neutrinos are on the same order, namely, $M^{}_1 \approx M^{}_2 = M$, such that both of them can be integrated out together at the energy scale $\mu = M$. Below the decoupling scale, the neutrino mass operator is then given by
\begin{eqnarray}
{\cal O}^{}_\nu = - \frac{1}{2} \kappa^*_{\alpha \beta} \widehat{L}^{}_\alpha \widehat{H}^{}_{\rm u} \widehat{L}^{}_\beta \widehat{H}^{}_{\rm u} \; ,
%     (4)
\label{eq:Onu}
\end{eqnarray}
and it is related to the effective neutrino mass matrix as $M^{}_\nu = \kappa v^2 \sin^2\beta$ after the electroweak gauge symmetry breaking.

Though the two-loop RG equations of physical parameters in both the full seesaw model and the corresponding effective theories have been known for a long time~\cite{Antusch:2002ek}, the one-loop matching conditions have been derived recently~\cite{Antusch:2015pda}. To calculate the smallest neutrino mass, we collect the relevant results from Refs.~\cite{Antusch:2015pda} and \cite{Antusch:2002ek} as below.
\begin{itemize}
\item At the decoupling scale $\mu = M$, the matching condition for the neutrino mass parameter is found to be~\cite{Antusch:2015pda}
\begin{eqnarray}
\kappa = \Delta \kappa^{}_0 + \Delta \kappa^{}_1 \; ,
%     (5)
\label{eq:match}
\end{eqnarray}
with the tree-level contribution
\begin{eqnarray}
\Delta \kappa^{}_0 = Y^{}_\nu M^{-1}_{\rm R} Y^{\rm T}_\nu \; ,
%     (6)
\label{eq:tree}
\end{eqnarray}
and the one-loop correction
\begin{eqnarray}
\Delta \kappa^{}_1 = -\frac{1}{32\pi^2} \left[ \Delta \kappa^{}_0 {\cal H}^{}_\nu + {\cal H}^{\rm T}_\nu \Delta \kappa^{}_0 + 2\, {\rm Tr}\left({\cal H}^{}_\nu\right) \Delta \kappa^{}_0 \right] \; ,
%    (7)
\label{eq:loop}
\end{eqnarray}
with ${\cal H}^{}_\nu \equiv Y^{}_\nu Y^\dagger_\nu$. From Eqs.~(\ref{eq:match})-(\ref{eq:loop}), one can observe that although $\Delta \kappa^{}_0$ is of rank two in the MSM, the addition of the one-loop contribution may increase the rank of the overall neutrino mass parameter $\kappa$ from two to three, as we shall show in the next section.

\item Below the decoupling scale $\mu < M$, the two-loop RG equation of the neutrino mass parameter $\kappa$ in the effective theory can be found in Ref.~\cite{Antusch:2002ek}:
\begin{eqnarray}
16\pi^2 \frac{{\rm d} \kappa}{{\rm d}t} = \alpha^{}_\kappa \kappa +  X^{\rm T}_\kappa \kappa + \kappa X^{}_\kappa \; ,
%     (8)
\label{eq:kappa}
\end{eqnarray}
where $t\equiv \ln(\mu/\Lambda^{}_{\rm F})$, and $\alpha^{}_\kappa = \alpha^{[1]}_\kappa + \alpha^{[2]}_\kappa$ and $X^{}_\kappa = X^{[1]}_\kappa + X^{[2]}_\kappa$ have been written in terms of the separated one-loop and two-loop contributions, namely,
\begin{eqnarray}
\alpha^{[1]}_\kappa &=& -\frac{6}{5} g^2_1 - 6 g^2_2 + 6\, {\rm Tr}\left({\cal H}^{}_{\rm u}\right) \; , \nonumber \\
\alpha^{[2]}_\kappa &=&  \frac{1}{16\pi^2} \left[ - 6\, {\rm Tr}\left({\cal H}^{}_{\rm d} {\cal H}^{}_{\rm u}\right) - 18\, {\rm Tr} \left( {\cal H}^2_{\rm u}\right) + \frac{8}{5} g^2_1\, {\rm Tr} \left( {\cal H}^{}_{\rm u} \right) + 32 g^2_3 \, {\rm Tr} \left( {\cal H}^{}_{\rm u} \right) \right. \nonumber \\
&~& \left. + \frac{207}{25} g^4_1 + \frac{18}{5} g^2_1 g^2_2 + 15 g^4_2 \right] \; ;
%     (9)
\label{eq:alphak}
\end{eqnarray}
and
\begin{eqnarray}
X^{[1]}_\kappa &=& {\cal H}^{}_l \; , \nonumber \\
X^{[2]}_\kappa &=& \frac{1}{16\pi^2} \left\{ -2\,{\cal H}^2_l + \left[ \frac{6}{5} g^2_1 - {\rm Tr}\left({\cal H}^{}_l\right) -3\, {\rm Tr}\left({\cal H}^{}_{\rm d}\right) \right] {\cal H}^{}_l\right\} \; ,
%     (10)
\label{eq:Xk}
\end{eqnarray}
where ${\cal H}^{}_x \equiv Y^{}_x Y^\dagger_x$ (for $x= {\rm u}, {\rm d}, l$) have been defined. As has already been noted in Ref.~\cite{Davidson:2006tg}, in the flavor basis where the charged-lepton Yukawa coupling matrix $Y^{}_l = {\rm Diag}\{y^{}_e, y^{}_\mu, y^{}_\tau\}$ is diagonal, the RG evolution of $\kappa$ cannot change its rank even at the two-loop level. Therefore, the supersymmetry (SUSY) breaking and the contributions from the soft $A$-terms have been introduced in Ref.~\cite{Davidson:2006tg} to generate the smallest neutrino mass $m^{}_{\rm L} \sim 10^{-10}~{\rm eV} \cdot (\tan \beta/10)^4$. The possible enhancement due to a large value of $\tan\beta$ is evident.
\end{itemize}

Since the one-loop matching condition in Eq.~(\ref{eq:match}) is available, one can assume that the SUSY is preserved until its breaking scale $\Lambda^{}_{\rm SUSY}$, which may be much lower than the typical seesaw scale $\Lambda^{}_{\rm SS} \sim 10^{14}~{\rm GeV}$, and estimate the smallest neutrino mass $m^{}_{\rm L}$ at the decoupling scale $\mu = M \sim \Lambda^{}_{\rm SS}$. Then, the two-loop RG equations can be implemented to run neutrino masses and other mixing parameters from the decoupling scale $\Lambda^{}_{\rm SS}$ to the SUSY breaking scale $\Lambda^{}_{\rm SUSY}$. Finally, the neutrino masses at the Fermi scale $\Lambda^{}_{\rm F}$ can be derived by incorporating the threshold effects due to the decoupling of superparticles and solving the RG equations between the SUSY breaking scale $\Lambda^{}_{\rm SUSY}$ and the Fermi scale $\Lambda^{}_{\rm F}$. However, this last step would very much depend on the supersymmetric particle spectrum and the symmetry breaking scheme, which are beyond the scope of the present work. In the following, we focus mainly on the neutrino mass parameter $\kappa$ at the decoupling scale $\mu = \Lambda^{}_{\rm SS}$ by taking into account the one-loop matching, and its RG running between the decoupling scale $\Lambda^{}_{\rm SS}$ and the SUSY breaking scale $\Lambda^{}_{\rm SUSY}$.

\section{Smallest Neutrino Mass}\label{sec:mass}

As mentioned in the last section, although the smallest neutrino mass in the MSM is vanishing at the tree level, the one-loop matching condition may increase the rank of the neutrino mass parameter $\kappa$ already at the decoupling scale $\mu = M$. Without loss of generality, we work in the flavor basis where the charged-lepton Yukawa coupling matrix is diagonal $Y^{}_l = {\rm Diag}\{y^{}_e, y^{}_\mu, y^{}_\tau\}$, and adopt the Casas-Ibarra parametrization of the Dirac neutrino Yukawa coupling matrix~\cite{Casas:2001sr}
\begin{eqnarray}
Y^{}_\nu = \frac{1}{v\sin\beta} U \sqrt{\widehat{M}^{}_\nu} R \sqrt{M^{}_{\rm R}} \; ,
%     (11)
\label{eq:Ynu}
\end{eqnarray}
where $M^{}_{\rm R} = {\rm Diag}\{M^{}_1, M^{}_2\}$ and
\begin{eqnarray}
\widehat{M}^{}_\nu = \left( \begin{matrix} 0 & 0 & 0 \cr 0 & m^{}_2 & 0 \cr 0 & 0 & m^{}_3 \end{matrix} \right) \; , \quad R = \left( \begin{matrix} 0 & 0 \cr \cos z & \sin z \cr - \sin z & \cos z \end{matrix} \right) \; ,
%     (12)
\label{eq:NO}
\end{eqnarray}
in the NO case with $m^{}_1 = 0$; or
\begin{eqnarray}
\widehat{M}^{}_\nu = \left( \begin{matrix} m^{}_1 & 0 & 0 \cr 0 & m^{}_2 & 0 \cr 0 & 0 & 0 \end{matrix} \right) \; , \quad R = \left( \begin{matrix} \cos z & \sin z \cr - \sin z & \cos z \cr  0 & 0 \end{matrix} \right) \; ,
%     (13)
\label{eq:IO}
\end{eqnarray}
in the IO case with $m^{}_3 = 0$. The $3\times 2$ matrix $R$ with $z$ being a complex number is orthogonal in the sense that $R^{\rm T} R = {\rm Diag}\{1, 1\}$ and $R R^{\rm T} = {\rm Diag}\{0, 1, 1\}$ in the NO case or $R R^{\rm T} = {\rm Diag}\{1, 1, 0\}$ in the IO case. Furthermore, we take the standard parametrization of the $3\times 3$ unitary matrix $U$~\cite{PDG2020}, where three rotation angles $\{\theta^{}_{12}, \theta^{}_{13}, \theta^{}_{23}\}$ and three CP-violating phases $\{\delta, \rho, \sigma\}$ are in general involved, i.e.,
\begin{eqnarray}
U = \left( \begin{matrix} c^{}_{13} c^{}_{12} & c^{}_{13} s^{}_{12} & s^{}_{13} e^{-{\rm i}\delta} \cr -s^{}_{12} c^{}_{23} - c^{}_{12} s^{}_{13} s^{}_{23} e^{{\rm i}\delta} & c^{}_{12} c^{}_{23} - s^{}_{12} s^{}_{13} s^{}_{23} e^{{\rm i}\delta} & c^{}_{13} s^{}_{23} \cr s^{}_{12} s^{}_{23} - c^{}_{12} s^{}_{13} c^{}_{23} e^{{\rm i}\delta} & -c^{}_{12} s^{}_{23} - s^{}_{12} s^{}_{13} c^{}_{23} e^{{\rm i}\delta} & c^{}_{13} c^{}_{23}\end{matrix} \right) \cdot \left( \begin{matrix} e^{{\rm i}\rho} & 0 & 0 \cr 0 & e^{{\rm i}\sigma} & 0 \cr 0 & 0 & 1 \end{matrix} \right) \; ,
%     (14)
\label{eq:U}
\end{eqnarray}
with $c^{}_{ij} \equiv \cos \theta^{}_{ij}$ and $s^{}_{ij} \equiv \sin \theta^{}_{ij}$ (for $ij = 12, 13, 23$). Notice that the CP-violating phase $\rho$ will be unphysical if the smallest neutrino mass is vanishing. Therefore, we can set $\rho = 0$ and take $\sigma$ to be the unique Majorana CP-violating phase in the case of either $m^{}_1 = 0$ or $m^{}_3 = 0$ at the energy scale $\mu = M$.

The neutrino mass parameter $\kappa$ at $\mu = M$ can be calculated by inserting $Y^{}_\nu$ in Eq.~(\ref{eq:Ynu}) into Eqs.~(\ref{eq:tree}) and (\ref{eq:loop}). For illustration, only the NO case is considered, whereas the IO case can be discussed in a similar way. First, we give the explicit expression of the Hermitian matrix
\begin{eqnarray}
{\cal H}^{}_\nu  = \frac{1}{v^2 \sin^2\beta} U \sqrt{\widehat{M}^{}_\nu} R M^{}_{\rm R} R^\dagger \sqrt{\widehat{M}^{}_\nu} U^\dagger = \frac{1}{v^2 \sin^2\beta} U \left( \begin{matrix} 0 & 0 & 0 \cr 0 & a & c \cr 0 & c^* & b \end{matrix} \right) U^\dagger \; ,
%     (15)
\label{eq:Hnu}
\end{eqnarray}
where
\begin{eqnarray}
a &\equiv& m^{}_2 \left(M^{}_1 |c^{}_z|^2 + M^{}_2 |s^{}_z|^2\right) \; ,
%     (16)
\label{eq:a} \\
b &\equiv& m^{}_3 \left(M^{}_1 |s^{}_z|^2 + M^{}_2 |c^{}_z|^2\right) \; ,
%     (17)
\label{eq:b} \\
c &\equiv& \sqrt{m^{}_2 m^{}_3} \left( M^{}_2 s^{}_z c^*_z - M^{}_1 s^*_z c^{}_z\right) \; ,
%     (18)
\label{eq:c}
\end{eqnarray}
with $c^{}_z \equiv \cos z$ and $s^{}_z \equiv \sin z$ being defined. From Eq.~(\ref{eq:Hnu}), it is straightforward to derive ${\rm Tr} \left( {\cal H}^{}_\nu \right) = \left[ (m^{}_2 M^{}_1 + m^{}_3 M^{}_2) |c^{}_z|^2 + (m^{}_2 M^{}_2 + m^{}_3 M^{}_1) |s^{}_z|^2 \right]/(v\sin\beta)^2$. Additionally, we have
\begin{eqnarray}
\Delta \kappa^{}_0 = \frac{1}{v^2\sin^2\beta} U \left( \begin{matrix} 0 & 0 & 0 \cr 0 & m^{}_2 & 0 \cr 0 & 0 & m^{}_3 \end{matrix} \right) U^{\rm T} \; ,
%     (16)
\label{eq:tree2}
\end{eqnarray}
which is actually related to the tree-level neutrino mass matrix via $M^{}_\nu = \Delta \kappa^{}_0 v^2 \sin^2\beta = U \widehat{M}^{}_\nu U^{\rm T}$. Then, by using Eqs.~(\ref{eq:Hnu}) and (\ref{eq:tree2}), one can figure out $\Delta \kappa^{}_1$ in Eq.~(\ref{eq:loop}) and thus the overall neutrino mass parameter $\kappa$ in Eq.~(\ref{eq:match}), which fulfills the relation
\begin{eqnarray}
U^\dagger \kappa U^* &=& \frac{1}{v^2 \sin^2\beta} \left[ 1 - \frac{(m^{}_2 M^{}_1 + m^{}_3 M^{}_2) |c^{}_z|^2 + (m^{}_2 M^{}_2 + m^{}_3 M^{}_1) |s^{}_z|^2}{16\pi^2 v^2 \sin^2\beta}\right] \widehat{M}^{}_\nu \nonumber \\
&~& -\frac{1}{32\pi^2 v^4 \sin^4\beta} \left[ \widehat{M}^{}_\nu U^{\rm T} U \left( \begin{matrix} 0 & 0 & 0 \cr 0 & a & c \cr 0 & c^* & b \end{matrix} \right) U^\dagger U^* + U^\dagger U^* \left( \begin{matrix} 0 & 0 & 0 \cr 0 & a & c^* \cr 0 & c & b \end{matrix} \right) U^{\rm T} U \widehat{M}^{}_\nu \right] \; ,
%     (20)
\label{eq:kappa2}
\end{eqnarray}
where one can observe that the symmetric unitary matrix ${\cal Z} \equiv U^{\rm T} U$ plays an important role in increasing the rank of $\kappa$. More explicitly, we get
\begin{eqnarray}
{\cal Z} = \left( \begin{matrix} 1 + 2{\rm i} c^2_{12} s^2_{13} s^{}_\delta e^{{\rm i}\delta} & 2{\rm i} s^{}_{12} c^{}_{12} s^2_{13} s^{}_\delta e^{{\rm i}(\delta + \sigma)} & -2{\rm i} c^{}_{12} s^{}_{13} c^{}_{13} s^{}_\delta \cr 2{\rm i} s^{}_{12} c^{}_{12} s^2_{13} s^{}_\delta e^{{\rm i}(\delta + \sigma)} &  \left(1 + 2{\rm i} s^2_{12} s^2_{13} s^{}_\delta e^{{\rm i}\delta}\right) e^{2{\rm i}\sigma} & -2{\rm i} s^{}_{12} s^{}_{13} c^{}_{13} s^{}_\delta e^{{\rm i}\sigma} \cr -2{\rm i} c^{}_{12} s^{}_{13} c^{}_{13} s^{}_\delta & -2{\rm i} s^{}_{12} s^{}_{13} c^{}_{13} s^{}_\delta e^{{\rm i}\sigma} & 1 - 2{\rm i} s^2_{13} s^{}_\delta e^{-{\rm i}\delta} \end{matrix} \right) \; ,
%     (21)
\label{eq:Z}
\end{eqnarray}
which will be reduced to the $3\times 3$ identity matrix in the absence of CP-violating phases (namely, $\delta = 0$ and $\sigma = 0$). If the Dirac CP-violating phase $\delta$ were trivial, the matrix on the left-hand side of Eq.~(\ref{eq:kappa2}) would be of rank two, leading to a vanishing neutrino mass $m^{}_1 = 0$. Therefore, we have to require $\delta$ in the unitary matrix $U$ to be nontrivial at the decoupling scale. Defining the symmetric matrix $\kappa^\prime \equiv U^\dagger \kappa U^* - \widehat{M}^{}_\nu/(v\sin\beta)^2$ stemming from one-loop matching, we obtain its six independent matrix elements
\begin{eqnarray}
\kappa^\prime_{11} &=& 0 \; , \nonumber \\
\kappa^\prime_{22} &=& -\frac{m^{}_2}{16\pi^2 v^4 \sin^4\beta} \left[a (1 + |{\cal Z}^{}_{22}|^2) + b (1 + |{\cal Z}^{}_{23}|^2) + 2 {\rm Re} \left( c {\cal Z}^{}_{22} {\cal Z}^*_{23} \right) \right] \; , \nonumber \\
\kappa^\prime_{33} &=& -\frac{m^{}_3}{16\pi^2 v^4 \sin^4\beta} \left[a (1 + |{\cal Z}^{}_{23}|^2) + b (1 + |{\cal Z}^{}_{33}|^2) + 2 {\rm Re} \left( c {\cal Z}^{}_{23} {\cal Z}^*_{33} \right) \right] \; ;
%     (22)
\label{eq:kpdiag}
\end{eqnarray}
and
\begin{eqnarray}
\kappa^\prime_{12} &=& -\frac{m^{}_2}{32\pi^2 v^4 \sin^4\beta} \left[ \left( a {\cal Z}^{}_{22} + c^* {\cal Z}^{}_{23}\right) {\cal Z}^*_{12} + \left( c {\cal Z}^{}_{22} + b {\cal Z}^{}_{23}\right) {\cal Z}^*_{13} \right] \; , \nonumber \\
\kappa^\prime_{13} &=& -\frac{m^{}_3}{32\pi^2 v^4 \sin^4\beta} \left[ \left( a {\cal Z}^{}_{23} + c^* {\cal Z}^{}_{33}\right) {\cal Z}^*_{12} + \left( c {\cal Z}^{}_{23} + b {\cal Z}^{}_{33}\right) {\cal Z}^*_{13} \right] \; , \nonumber \\
\kappa^\prime_{23} &=& -\frac{1}{32\pi^2 v^4 \sin^4\beta} \left[ (m^{}_2 + m^{}_3) {\rm Re} \left( a {\cal Z}^{}_{22} {\cal Z}^*_{23} + c {\cal Z}^{}_{22} {\cal Z}^*_{33} + c^* |{\cal Z}^{}_{23}|^2 + b {\cal Z}^{}_{23} {\cal Z}^*_{33}\right) \right. \nonumber \\
&~& \hspace{2.45cm} \left. + {\rm i} (m^{}_2 - m^{}_3) {\rm Im} \left( a {\cal Z}^{}_{22} {\cal Z}^*_{23} + c {\cal Z}^{}_{22} {\cal Z}^*_{33} + c^* |{\cal Z}^{}_{23}|^2 + b {\cal Z}^{}_{23} {\cal Z}^*_{33} \right) \right] \; . \qquad
%     (23)
\label{eq:kpoff}
\end{eqnarray}
Some comments on the effective neutrino mass parameter $\kappa = U [ \widehat{M}^{}_\nu/(v\sin\beta)^2 + \kappa^\prime ] U^{\rm T}$ with $\kappa^\prime$ given in Eqs.~(\ref{eq:kpdiag}) and (\ref{eq:kpoff}) are in order.
\begin{itemize}
\item First, when the one-loop contribution $\Delta \kappa^{}_1$ in Eq.~(\ref{eq:loop}) is included, the nonzero off-diagonal elements $\kappa^\prime_{12}$ and $\kappa^\prime_{13}$ of the matrix $\kappa^\prime$ are generated and they may lead to a nonvanishing neutrino mass $m^{}_1$. Notice that $\kappa^\prime_{11}$ itself remains to be vanishing, so the smallest neutrino mass $m^{}_1$ can be obtained only after diagonalizing the symmetric matrix $\kappa^\prime$, which also renders the final mixing matrix to be different from $U$. Once the smallest neutrino mass becomes nonzero, the associated Majorana CP-violating phase $\rho$ will be physical~\cite{Xing:2020ezi}.

\item Second, it is worthwhile to stress that the off-diagonal elements $\kappa^\prime_{12}$ and $\kappa^\prime_{13}$ in Eq.~(\ref{eq:kpoff}) are proportional to $\sin\delta$, so a nontrivial Dirac-type CP-violating phase $\delta$ is crucially important for generating the smallest neutrino mass. For instance, if $\delta = 0$ or $\pi$ holds, then we have
    \begin{eqnarray}
    {\cal Z} = {\rm Diag}\{1, e^{2{\rm i}\sigma}, 1\} \; ,
    %     (24)
    \label{eq:Ztrivial}
    \end{eqnarray}
    from Eq.~(\ref{eq:Z}). In this case, the right-hand side of Eq.~(\ref{eq:kappa2}) will be reduced to a $3\times 3$ matrix with vanishing elements in the first row and first column, indicating $m^{}_1 = 0$.
\end{itemize}

As one can see from previous observations, the smallest neutrino mass depends on the general form of $\kappa^\prime$, in which a number of free high-energy parameters in the MSM are involved. To make an order-of-magnitude estimation of $m^{}_{\rm L}$, we take the following assumptions and approximations. First, since $\sin^2 \theta^{}_{13} \approx 0.022$ is observed at the Fermi scale and the RG running of $\theta^{}_{13}$ is insignificant with a small value of $\tan \beta = 10$, one may ignore all the terms of ${\cal O}(s^2_{13})$ in the symmetric unitary matrix ${\cal Z}$ in Eq.~(\ref{eq:Z}), namely,
\begin{eqnarray}
{\cal Z} \approx \left( \begin{matrix} 1 & 0 & 0 \cr 0 & e^{2{\rm i}\sigma} & 0 \cr 0 & 0 & 1\end{matrix} \right) - 2{\rm i} s^{}_{13} c^{}_{13} s^{}_\delta \left( \begin{matrix} 0 & 0 & c^{}_{12} \cr 0 & 0 & s^{}_{12} e^{{\rm i}\sigma} \cr c^{}_{12} & s^{}_{12} e^{{\rm i}\sigma} & 0 \end{matrix} \right) \; .
%     (25)
\label{eq:Zapp}
\end{eqnarray}
Second, we further take $M^{}_1 = M^{}_2 = M$ and ${\rm Im}z = 0$ at the decoupling scale $\mu = M$, and thus $a = m^{}_2 M$, $b = m^{}_3 M$ and $c = 0$ from Eqs.~(\ref{eq:a})-(\ref{eq:c}) can be obtained. With these assumptions, one can easily verify that the matrix elements in Eqs.~(\ref{eq:kpdiag}) and (\ref{eq:kpoff}) approximate to
\begin{eqnarray}
\kappa^\prime_{11} &=& 0 \; , \nonumber \\
\kappa^\prime_{22} &\approx& - \frac{m^{}_2 (2m^{}_2 + m^{}_3) M }{16\pi^2 v^4 \sin^4\beta}\; , \nonumber \\
\kappa^\prime_{33} &\approx& - \frac{m^{}_3 (m^{}_2 + 2 m^{}_3) M }{16\pi^2 v^4 \sin^4\beta}\; ;
%     (26)
\label{eq:kpdiagapp}
\end{eqnarray}
and
\begin{eqnarray}
\kappa^\prime_{12} &\approx& 0\; , \nonumber \\
\kappa^\prime_{13} &\approx& - \frac{{\rm i} m^2_3 M }{16\pi^2 v^4 \sin^4\beta} c^{}_{12} s^{}_{13} c^{}_{13} s^{}_\delta \; , \nonumber \\
\kappa^\prime_{23} &\approx& + \frac{\left[s^{}_\sigma (m^2_2 - m^2_3) - {\rm i}c^{}_\sigma (m^{}_2 - m^{}_3)^2 \right] M }{16\pi^2 v^4 \sin^4\beta} s^{}_{12} s^{}_{13} c^{}_{13} s^{}_\delta \; ,
%     (27)
\label{eq:kpoffapp}
\end{eqnarray}
where the higher-order terms ${\cal O}(s^2_{13})$ have been omitted. Now it is clear that the smallest neutrino mass with the one-loop matching conditions can be calculated by diagonalizing the matrix $\kappa^\prime$. At the leading order, the smallest eigenvalue of the effective neutrino mass matrix $M^{}_\nu = \kappa v^2 \sin^2 \beta$ is found to be
\begin{eqnarray}
m^\prime_1 \approx m^{}_3 c^2_{12} s^2_{13} c^2_{13} s^2_\delta \left( \frac{m^{}_3 M}{16\pi^2 v^2 \sin^2\beta}\right)^2 \; .
%     (28)
\label{eq:m1prime}
\end{eqnarray}
If the numerical values $\theta^{}_{12} \approx 33.44^\circ$, $\theta^{}_{13} \approx 8.57^\circ$, $\delta \approx 195^\circ$ and $m^{}_3 = \sqrt{\Delta m^2_{31}} \approx 50.1~{\rm meV}$ from neutrino oscillation data~\cite{Esteban:2020cvm}, together with $\tan\beta = 10$ and $M = 10^{14}~{\rm GeV}$, are inserted into Eq.~(\ref{eq:m1prime}), we get
\begin{eqnarray}
m^\prime_1 \approx 5.6\times 10^{-11}~{\rm eV} \cdot \left(\frac{\sin^2\delta}{0.067}\right) \cdot \left( \frac{m^{}_3}{50.1~{\rm meV}}\right)^3 \cdot \left( \frac{M}{10^{14}~{\rm GeV}} \right)^2 \; .
%     (29)
\label{eq:m1primenum}
\end{eqnarray}
It should be noticed that $\sin^2\beta = \tan^2\beta/(1 + \tan^2\beta) \approx 1$ is actually insensitive to the value of $\tan\beta$ as long as $\tan^2\beta \gg 1$ holds. The estimation of $m^\prime_1$ in Eq.~(\ref{eq:m1primenum}) is rather conservative in two aspects. First, the size of the Dirac Yukawa coupling ${\cal O}(Y^{}_\nu) \sim m^{}_3 M/(v\sin\beta)^2 \approx 0.167$ is small for the adopted values $m^{}_3 = 50.1~{\rm meV}$ and $M = 10^{14}~{\rm GeV}$. If $M = 10^{15}~{\rm GeV}$ is chosen, the perturbativity constraint ${\cal O}(Y^{}_\nu) < \sqrt{4\pi}$ is still satisfied while $m^\prime_1$ will be enhanced by two orders of magnitude, i.e., $m^\prime_1 \sim 5.6\times 10^{-9}~{\rm eV}$. Second, the CP-violating phase $\delta$ could even be maximal at the seesaw scale such that $\sin^2\delta = 1$, leading to the enhancement of $m^\prime_1$ by another order of magnitude. Therefore, it is reasonable to conclude that the smallest neutrino mass at the decoupling scale can reach $m^\prime_1 \sim 10^{-8}~{\rm eV}$.

After fixing the neutrino mass parameter $\kappa$ by taking account of the one-loop matching at the decoupling scale $\mu = M = \Lambda^{}_{\rm SS}$, one has to solve the two-loop RG equation of $\kappa$ in Eq.~(\ref{eq:kappa}) and derive the smallest neutrino mass at the SUSY breaking scale $\Lambda^{}_{\rm SUSY}$. Since the two-loop beta functions in Eqs.~(\ref{eq:alphak}) and (\ref{eq:Xk}) are always suppressed by the loop factor $1/(16\pi^2)$ when compared to the one-loop beta functions, it is safe to examine the RG running effects on neutrino masses by considering only the one-loop beta functions. In the flavor basis where the charged-lepton Yukawa coupling matrix is diagonal, we have $Y^{}_l = {\rm Diag}\{y^{}_e, y^{}_\mu, y^{}_\tau\}$ and ${\cal H}^{}_l = {\rm Diag}\{y^2_e, y^2_\mu, y^2_\tau\}$. The solution to the RG equation of $\kappa$ in Eq.~(\ref{eq:kappa}) can be easily derived
\begin{eqnarray}
\kappa(\Lambda^{}_{\rm SUSY}) = I^{}_0 \cdot T^{}_l \cdot \kappa(\Lambda^{}_{\rm SS}) \cdot T^{}_l \; ,
%     (30)
\label{eq:kappaSUSY}
\end{eqnarray}
where $T^{}_l \equiv {\rm Diag}\{I^{}_e, I^{}_\mu, I^{}_\tau\}$ has been defined and the relevant evolution functions read
\begin{eqnarray}
I^{}_0 &=& \exp\left[ - \frac{1}{16\pi^2} \int^{\ln(\Lambda^{}_{\rm SS}/\Lambda^{}_{\rm SUSY})}_0 \alpha^{}_\kappa(t) {\rm d}t \right] \; ,
%     (31)
\label{eq:I0} \\
I^{}_\alpha &=& \exp\left[ - \frac{1}{16\pi^2} \int^{\ln(\Lambda^{}_{\rm SS}/\Lambda^{}_{\rm SUSY})}_0 y^2_\alpha(t) {\rm d}t \right]\; ,
%     (32)
\label{eq:Ialpha}
\end{eqnarray}
where the charged-lepton Yukawa couplings are given by $y^2_\alpha = m^2_\alpha (1 + \tan^2\beta) /v^2$ (for $\alpha = e, \mu, \tau$). In principle, the charged-lepton Yukawa couplings $y^{}_\alpha$ can be greatly enhanced for a large value of $\tan\beta$, implying a significant running effect on the lepton flavor mixing parameters~\cite{Antusch:2003kp}. However, the dominant effect on the running of neutrino masses arises from the evolution function $I^{}_0$, in which the integrand is $\alpha^{}_\kappa \approx - 6 g^2_1/5 - 6g^2_2 + 6 y^2_t$ with $y^{}_t \approx 1$ being the top-quark Yukawa coupling constant. Numerically, for $\Lambda^{}_{\rm SS} = 10^{14}~{\rm GeV}$ and $\Lambda^{}_{\rm SUSY} = 10^4~{\rm GeV}$, we find that $I^{}_0 \approx 0.8$ for $\tan\beta = 10$, implying that the smallest neutrino mass $m^{}_1(\Lambda^{}_{\rm SUSY})$ will be reduced by about $20\%$ when compared to the value $m^{}_1(\Lambda^{}_{\rm SS})$~\cite{Mei:2003gn}. Consequently, we obtain the smallest neutrino mass
\begin{eqnarray}
m^{}_1(\Lambda^{}_{\rm SUSY}) \approx I^{}_0 m^\prime_1(\Lambda^{}_{\rm SS}) \approx 6.7\times 10^{-10}~{\rm eV} \cdot \left( \frac{m^{}_3}{50.1~{\rm meV}}\right)^3 \cdot \left( \frac{M}{10^{14}~{\rm GeV}} \right)^2 \; ,
%     (33)
\label{eq:m1final}
\end{eqnarray}
where $\sin^2\delta = 1$ and $I^{}_0 \approx 0.8$ have been used. Though there will be additional corrections from the SUSY threshold effects~\cite{Antusch:2008tf} and the RG running in the effective theory below the SUSY breaking scale~\cite{Xing:2007fb}, we expect that the smallest neutrino mass at the Fermi scale $m^{}_1(\Lambda^{}_{\rm F})$ should be on the same order of that derived in Eq.~(\ref{eq:m1final}). 

In consideration of the uncertainties on the choice of the seesaw scale $\Lambda^{}_{\rm SS}$ and the input values of relevant parameters at $\Lambda^{}_{\rm SS}$, we may finally estimate the lightest neutrino mass at the Fermi scale $\Lambda^{}_{\rm F}$ to be in the range $m^{}_1 \in [10^{-10}, 10^{-8}]~{\rm eV}$. The lower bound of this range corresponds to the case with the choice of $\Lambda^{}_{\rm SS} = 10^{14}~{\rm GeV}$ and the assumption that the relevant parameters at $\Lambda^{}_{\rm SS}$ take the corresponding best-fit values at the low-energy scale. On the other hand, the upper bound is obtained by taking $\Lambda^{}_{\rm SS} = 10^{15}~{\rm GeV}$ and assuming the best-fit values of neutrino oscillation parameters at the low-energy scale, except that the maximum of the Dirac-type CP-violating phase (i.e., $\sin^2\delta = 1$) is now adopted. More accurate calculations of three neutrino masses and flavor mixing parameters in the MSM can be done numerically by scanning the full parameter space.

\section{Summary}\label{sec:sum}

In the present work, we point out that the one-loop matching between the full theory and the effective theory can generate a nonzero mass of the lightest neutrino in the minimal seesaw model. In the previous works, the two-loop RG equation of the effective neutrino mass parameter $\kappa$ has been implemented in the SM framework, whereas the SUSY breaking has to be considered in the MSSM framework~\cite{Davidson:2006tg}.

In the case of normal neutrino mass ordering with $m^{}_1 = 0$ at the seesaw scale $\Lambda^{}_{\rm SS} = 10^{14}~{\rm GeV}$, we explicitly demonstrate that the one-loop matching condition for the neutrino mass parameter $\kappa = \Delta \kappa^{}_0 + \Delta \kappa^{}_1$, where $\Delta \kappa^{}_0$ and $\Delta \kappa^{}_1$ denote respectively the tree- and one-loop contributions, can lead to a nonzero neutrino mass at the Fermi scale $\Lambda^{}_{\rm F} = 91.2~{\rm GeV}$. By using the one-loop matching condition available from Ref.~\cite{Antusch:2015pda} in the supersymmetric version of the minimal seesaw model, we find that the smallest neutrino mass $m^{}_1(\Lambda^{}_{\rm SS})$ could be as large as $10^{-8}~{\rm eV}$, depending on the parameters in the full seesaw model above the decoupling scale. In particular, when the Casas-Ibarra parametrization of the Dirac neutrino Yukawa coupling matrix $Y^{}_\nu = U (\widehat{M}^{}_\nu)^{1/2} R (M^{}_{\rm R})^{1/2}$ is adopted, the Dirac-type CP-violating phase $\delta$ in the unitary matrix $U$ must take a nontrivial value to guarantee a nonzero mass $m^{}_1 \propto \sin^2\delta$. The rotation angles in $U$ should also take nontrivial values, otherwise the Dirac CP-violating phase $\delta$ could have been eliminated. For some typical input at the seesaw scale, we obtain the smallest neutrino mass $m^{}_1(\Lambda^{}_{\rm F}) \in [10^{-10}, 10^{-8}]~{\rm eV}$ after taking account of the RG running effects on neutrino masses.

It is worth emphasizing that the smallest neutrino mass from one-loop matching depends very much on the flavor structure of the Dirac neutrino Yukawa coupling matrix $Y^{}_\nu$ and heavy Majorana neutrino masses $M^{}_i$ (for $i = 1, 2$). A complete study of all the allowed parameters in the full seesaw model and their impact on the generated neutrino masses and flavor mixing parameters at the Fermi scale is interesting and necessary. We also expect that the one-loop matching can generate a nonzero smallest neutrino mass in the minimal seesaw model in the SM framework. We shall come back to these issues in the near future.

\section*{Acknowledgements}

The author would like to thank Di Zhang for helpful discussions, and Prof. Zhi-zhong Xing for valuable suggestions. This work was supported in part by the National Natural Science Foundation of China under Grant No.~11775232 and No.~11835013, and by the CAS Center for Excellence in Particle Physics.

\end{document}